\documentclass[a4paper,12pt]{article}
\usepackage{graphicx} %use graph format
\usepackage{epstopdf}

\begin{document}

%%-move to normal A4-%%
\hoffset = -1truecm \voffset = -2truecm \baselineskip = 10 mm

\title{\bf The gluon condensation in high energy cosmic rays
}

\author{
{\bf Wei Zhu$^1$\footnote{Corresponding author, E-mail:
weizhu@mail.ecnu.edu.cn}}, {\bf Jiangshan Lan$^2$} and {\bf Jianhong
Ruan}$^1$
\\
\normalsize $^1$Department of Physics, East China Normal University,
Shanghai 200241, China \\
\normalsize $^2$Institute of Modern Physics, Chinese Academy of
Sciences, Lanzhou 730000,
 China\\
}

\date{}

\newpage

\maketitle

\vskip 3truecm

\begin{abstract}

    The gluon condensation (GC)-effects in high energy cosmic rays are
investigated. After a brief review of the GC, several examples
including gamma-, electron-, and positron-spectra in a broad
GeV$\sim $TeV region can be explained by the GC-effects. We find
that the GC may break the power-law of the cosmic ray spectra if the
energy of accelerated protons exceeds the GC-threshold. The GC is a
new phenomenon that is not yet known, it provides a new window to
understand the high energy cosmic ray spectra.

\end{abstract}

{\bf keywords}: Quantum Chromodynamics; Gluon condensation; Cosmic ray spectra

{\bf PACS numbers}:12.38.-t; 14.70.Dj; 14.20.Dh; 95.30.Cq

\newpage
\begin{center}
\section{Introduction}
\end{center}

    The majority of high energy particles in cosmic rays are protons.
Proton-proton (or nuclei) ($p-p(A)$) collisions are general events
in Universe. Gluons inside proton dominate the proton collisions at
high energy and their distributions obey the evolution equations
based on Quantum Chromodynamics (QCD). QCD analysis shows that the
evolution equations will become nonlinear due to the initial gluons
correlations at high energy and these will result in the chaotic
solution beginning at a threshold energy [1,2]. Most surprisingly,
the dramatic chaotic oscillations produce strong shadowing and
antishadowing effects, they converge gluons to a state at a critical
momentum [3]. This is the gluon condensation (GC) in proton. We will
give a brief review about its history in Sec. 2.

    The GC-effects should induce significant effects in the proton
processes, provided the GC-threshold $E_{p-p(A)}^{GC}$ enters the
observable high energy region. Unfortunately, the exact value of
$E_{p-p(A)}^{GC}$ can not been entirely determined in theory since
it relates to the unknown input conditions. On the other hand, we
have not observed the GC-effects until 13 TeV in $p-p$ collisions
and $5\sim 8$ TeV in $Pb-Pb$ collisions at the Large Hadron Collider
(LHC). Therefore, we turn to study the energy spectra of high energy
cosmic rays. Protons accelerated in the Universe may exceed
$E_{p-p(A)}^{GC}$ and cause the GC-effects in their collisions.

    The power-law form of energy spectrum is a general rule
of the cosmic ray spectra at high energy. It is described by a
straight line of the energy spectrum with a fixed index in a log-log
representation. This line may span over more than one order of
magnitude. The broken power-law relates to the extra sources of
cosmic rays or vent a new effect. We noticed that there are
different radiation mechanisms. Usually even one $\gamma$-ray
spectrum may arise the arguments of leptonic or hadronic
explanations. Therefore, we ask what is the GC-characteristic
spectrum? Can it distinguish the GC-effects from other phenomena? We
will discuss the above questions at Sec. 3.

     Following the results of Sec.3, we calculate
the gamma spectrum of a supernova remnant (SNR) Tycho [4] in Sec. 4.
We assume that the protons accelerated in this SNR may exceed
$E_{p-p(A)}^{GC}$ ($A$ indicates the nucleon number of a targeted
nucleus). Thus, the GC-effects dominate their energy spectra if
accelerated protons interact with the surrounding dense matter
inside the source. We find that our predicted sharp broken power-law
is consistent with the observed data. Following this example, a
several sharp broken power examples of the gamma ray originating
from SNRs and Active Galactic Nuclei (AGN) are discussed.

    An excess of the cosmic ray positron spectra at 10 GeV-1 TeV has been
reported by the Alpha Magnetic Spectrometer (AMS02) [5], which was
interpreted as the DM-signature. A similar excess structure in the
electron-positron spectrum at 300 GeV-700 GeV was early reported by
the Advanced Thin Ionization Calorimeter (ATIC) [6]. However, the
later finding was not confirmed by more accurate observations of the
the Fermi Large Area Telescope (Fermi-LAT) [7], the High Energy
Stereoscopic System (H.E.S.S.) [8,9], the Major Atmospheric Gamma
Imaging Cherenkov (MAGIC) [10] and the VERITAS [11]. The new data
show that a smooth broken power in the electron energy spectrum may
expand to a broader range rather than a narrow excess. We try to
explain these results using the GC-effects in Sec. 5. Finally, a
summary is given in Sec. 6.

\newpage
\begin{center}
\section{A brief review about the gluon condensation}
\end{center}
\begin{figure}
	\begin{center}
		\includegraphics[width=0.8\textwidth]{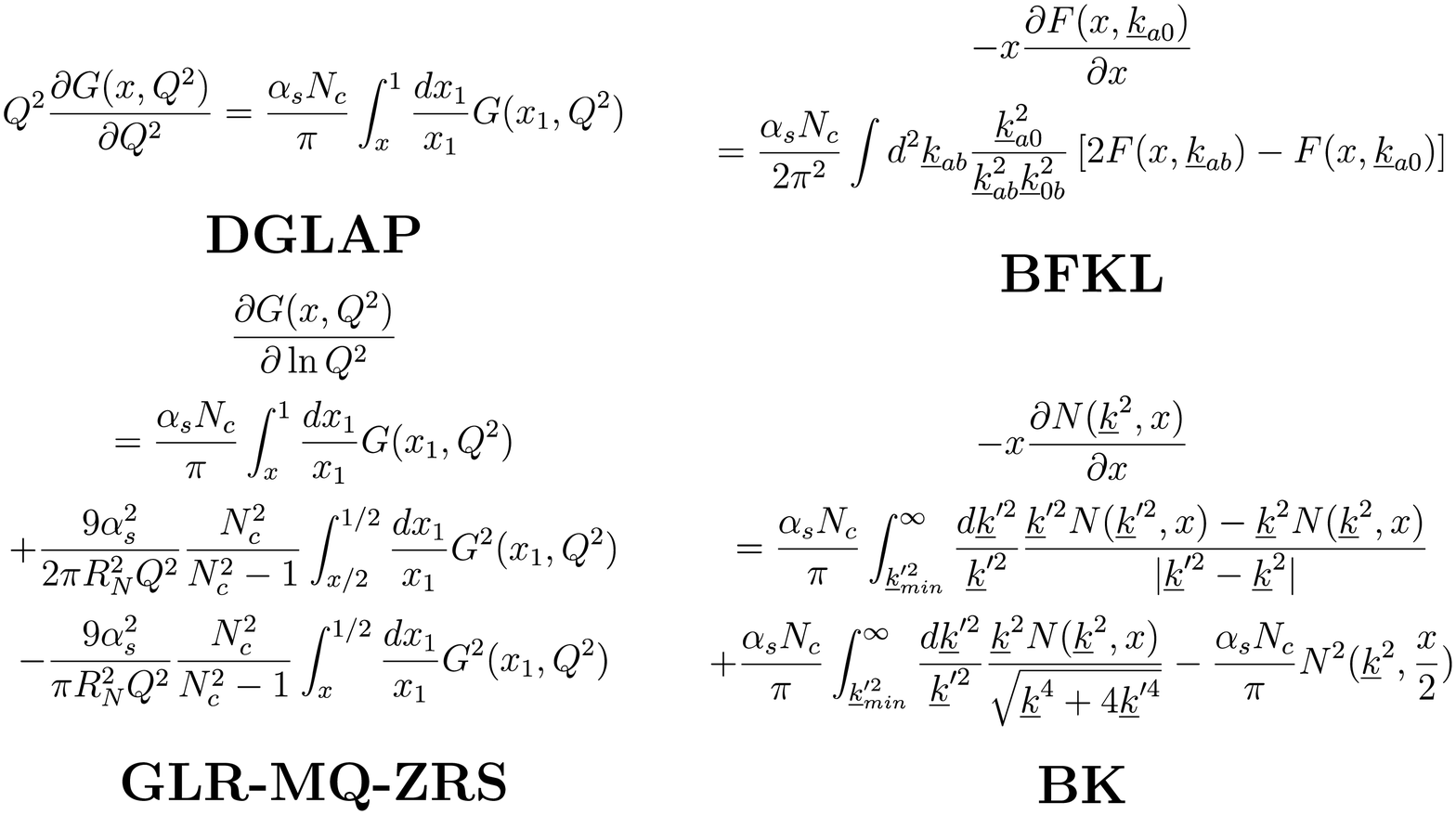} 
		\caption{A set of the QCD evolution equations for gluons
			at small $x$ and leading order approximation. Where the BK equation
			is taken in a full momentum space. $G(x,Q^2)$ and $F(x,
			\underline{k})$ are integrated and unintegrated gluon distributions.}\label{fig:1}
	\end{center}
\end{figure}
   
     The QCD evolution equations have different forms at different energy
ranges. At high energy, or small Bjorken variable $x$, gluon
distributions dominate the evolution processes (Fig. 1) according to
the linear DGLAP (Dokshitzer-Gribov-Lipatov-Altarelli-Parisi)
equation [12-14] and BFKL (Balitsky-Fadin-Kuraev-Lipatov) equation
[15-19]. They both predict that the gluon density in proton grows
with decreasing $x$ and it will cause the violation of the unitarity
of the scattering cross section. In consequence, a series of the
nonlinear evolution equations, for example, the GLR-MQ-ZRS
(Gribov-Levin-Ryskin, Mueller-Qiu, Zhu-Ruan-Shen) equation [20-24]
and BK (Balitsky-Kovchegov) equation[25,26] were proposed, in which
the corrections of gluon recombination are considered. An important
result of the non-linearization is that the BK equation, or its
generalization, the
Jalilian-Marian-Iancu-McLerran-Weigert-Leonidov-Kovner (JIMWLK)
equation [27-29],  predicts the unintegrated gluon distribution
$F(x,\underline{k})\rightarrow$ a constant when the transverse
momentum $\underline{k}$ of gluon is smaller than a characteristic
saturation momentum $Q_s(x)$. This saturation behavior implies a
balance between gluon splitting and fusion, however it is also
understood as the color glass condensate (CGC), where ``condensate"
implies that the maximum occupation number of gluons is $\sim
1/\alpha_s> 1$, although it lacks a characteristic sharp peak in the
momentum distribution (Fig. 2 (a-c)).
\begin{figure}
	\begin{center}
		\includegraphics[width=0.8\textwidth]{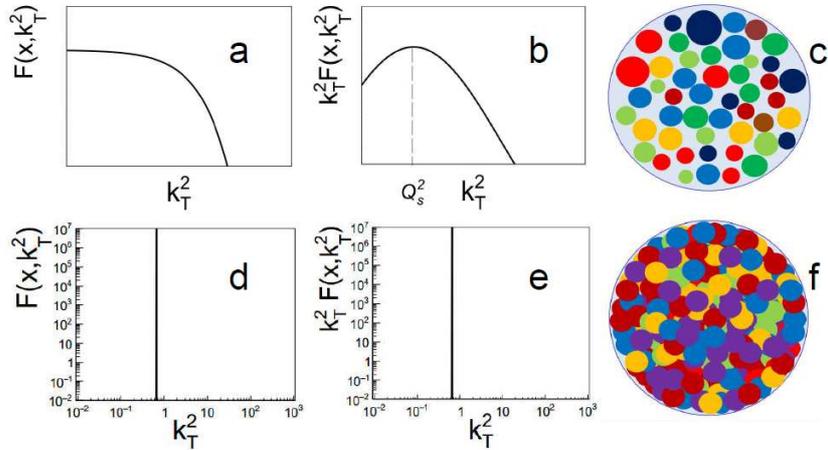} 
		\caption{Schematic diagrams (a-c) for the saturation
			solutions of the BK/JIMWLK equation; (d-f) for the condensation
			solutions of the ZSR equation, which is evolved with the GBW input.}\label{fig:2}
	\end{center}
\end{figure}

\begin{figure}
	\begin{center}
		\includegraphics[width=0.8\textwidth]{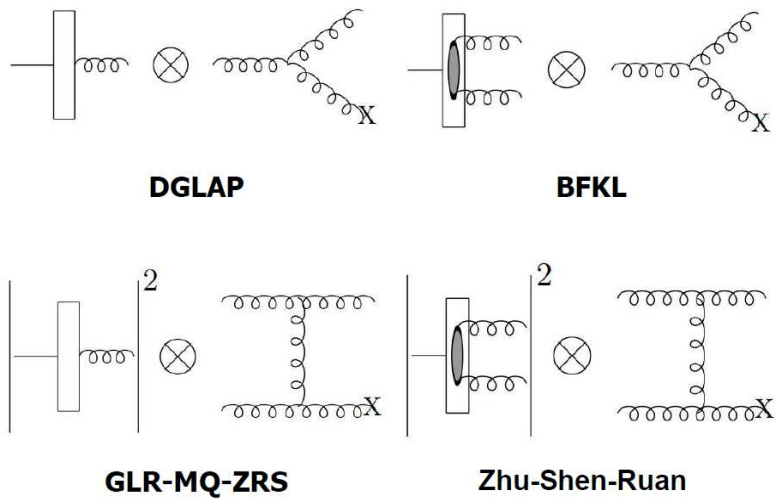} 
		\caption{ A set of consistent elemental amplitudes. Note
			that a complete evolution equation based on (d) should contain
			interference and virtual processes. These four evolution equations
			were unitary derived using the time ordered perturbative theory in
			[1,2].}\label{fig:3}
	\end{center}
\end{figure}

     A new nonlinear evolution equation based on Fig. 3
is derived by Zhu, Shen and Ruan (ZSR) in works [1,2]. According to
the standard quantum field theory, summing-up all possible
amplitudes including so-called virtual processes are necessary for
the regularization of the evolution equations. The derivation of the
nonlinear evolution equations based on Fig. 3 includes the
contributions from real 2-2, virtual 2-2, real 1-3 and virtual 1-3
amplitudes. A key point is how to calculate the above complicated
virtual diagrams? Fortunately, such a technique was established
using the time ordered perturbation theory (TOPT) in work [22]. This
TOPT-cutting rule was successfully used to recover the momentum
non-conservation in the GLR-MQ equation [23] and obtained the
support in serious of examples [30-35]. Using the TOPT cutting rule,
the resulting ZSR equation for the unintegrated gluon distribution
$F(x,\underline{k}^2)$ at the leading logarithmic $(LL(1/x))$
approximation is [2]

$$-x\frac{\partial F(x,\underline{k}^2)}{\partial x}$$
$$=\frac{3\alpha_{s}\underline{k}^2}{\pi}\int_{\underline{k}^2_0}^{\infty} \frac{d
\underline{k}'^2}{\underline{k}'^2}\left\{\frac{F(x,\underline{k}'^2)-F(x,\underline{k}^2)}
{\vert
\underline{k}'^2-\underline{k}^2\vert}+\frac{F(x,\underline{k}^2)}{\sqrt{\underline{k}^4+4\underline{k}'^4}}\right\}$$
$$-\frac{81}{16}\frac{\alpha_s^2}{\pi R^2_N}\int_{\underline{k}^2_0}^{\infty} \frac{d
\underline{k}'^2}{\underline{k}'^2}\left\{\frac{\underline{k}^2F^2(x,\underline{k}'^2)-\underline{k}'^2F^2(x,\underline{k}^2)}
{\underline{k}'^2\vert
\underline{k}'^2-\underline{k}^2\vert}+\frac{F^2(x,\underline{k}^2)}{\sqrt{\underline{k}^4+4\underline{k}'^4}}\right\},
\eqno(1)$$ where the first item on the right is the BFKL equation
part and the second one is nonlinear correction caused by the gluon
fusions. The singular structure in the linear and nonlinear
evolution kernels both corresponds to the random evolution in the
$\underline{k}$-space, where $\underline{k}^2-\underline{k}'^2$ may
cross over zero. This is a general property of the logarithmic
($1/x$) resummation, and remember, this is also a key point of our
following work.

       It is interest that the nonlinear ZSR evolution equation results in the chaotic
solution if the Bjorken variable $x$ goes beyond a critical point
$x_c$. This is a chaotic example in the QCD evolution equations,
although chaos is a popular natural phenomenon in nonlinear science.
Most surprisingly, the dramatic chaotic oscillations produce the
strong shadowing and antishadowing effects via the nonlinear terms
of Eq. (1), they converge gluons to a state with a critical momentum
(Figs. 2(d-f)). This is the gluon condensation (GC) in proton (see
Fig. 4(a), for comparison, we present the GBW model in Fig. 4(b),
which describes the CGC solution).

\begin{figure}
	\begin{center}
		\includegraphics[width=0.8\textwidth]{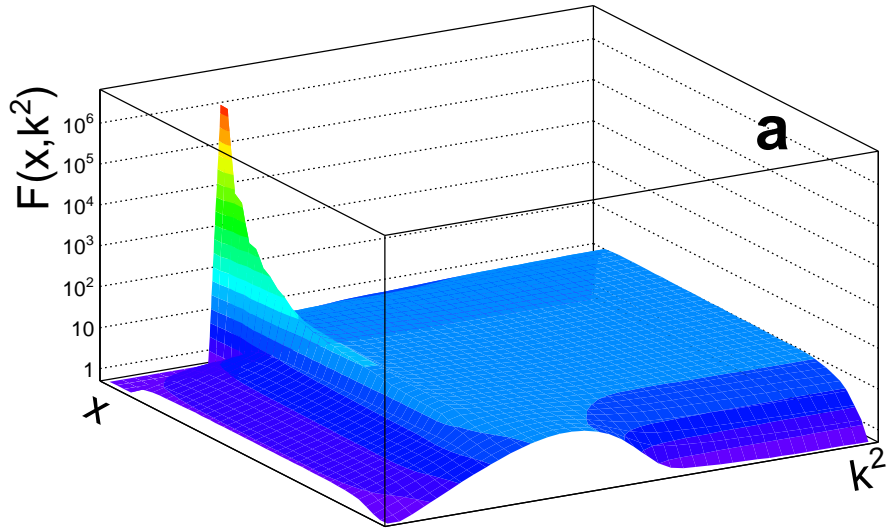} 
		\includegraphics[width=0.8\textwidth]{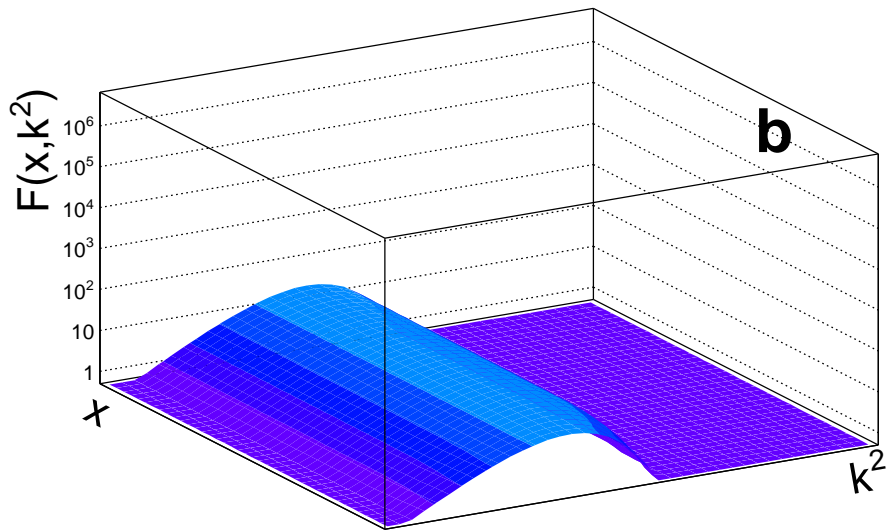} 
		\caption{(a) A GC-solution of the ZSR equation; (b) The
			GBW input distribution.}\label{fig:4}
	\end{center}
\end{figure}

    We emphasize that the above chaotic solution or the
convergence effect origins from the regularized nonlinear part in
Eq. (1), and it is a general structure of the logarithmic ($1/x$)
resummation even at higher ordered approximations. One can
understand this point as follows.

    (i) The gluon transverse momentum after the above mentioned
splitting process may either larger or smaller than that the
momentum of the primary gluon. As we know that the contributions of
evolution along random chain of gluon-transverse momentum
$\underline{k}$ are important at high energy. This point is always
emphasized by the classical linear BFKL evolution equation,
unfortunately, it was neglected, or is purposely avoided by other
nonlinear QCD evolution equations. Comparing with the BK-JIMWLK
equations, the ZSR equation (1) keeps the regularized nonlinear
terms and they have the derivative structure $\sim
\partial F(x,\underline{k}^2)/\partial \underline{k}^2$ and $\sim \partial
F^2(x,\underline{k}^2)/\partial\underline{k}^2$ in Eq. (1). These
terms may add a perturbation on the smooth curve
$F(x,\underline{k}^2)$ once $\vert\underline{k}\vert$ crosses over
$Q_s$. Thus, we have a series of independent perturbations in a
narrow $\vert\underline{k}\vert$ domain near $Q_s$ along evolution
to smaller $x$. In the linear BFKL equation, these perturbations are
independent and their effects are negligibly small. In this case the
solutions keep the smooth curves in ($x$, $\underline{k}^2$) space.
However, the nonlinear Eq. (1) may form chaos near $Q_s$. The
positive Lyapunov exponents of Eq. (1) support this suggestion.

    (ii) Note that once chaos is produced, the fast oscillations of the
gluon density produce both the negative and positive nonlinear
corrections to $\Delta F(x,\underline{k}^2)$ through the derivative
structure of Eq. (1). The former shadows the grownup of the gluon
density, while the later is the antishadowing effect, and it
increases rapidly the gluon density because it is a strong positive
feedback process. They will result in a pair of closer and more
stronger positive and negative corrections at a next evolution step,
where the positive correction continually put $F(x,\underline{k}^2)$
toward a bigger value, while the negative one suppress all remaining
distributions. Thus, we observed the gluons condensation at
($x_c,\underline{k}^2_c$) due to the extrusion of the shadowing and
antishadowing effects in the QCD evolution. (see [2,3] for details).

    Equation (1) is based on the leading QCD approximation, where
the higher order corrections are neglected. An important question
is: will the chaos effects disappear in the evolution equation after
considering higher order corrections? In other words, what is the
source of the GC-effects? and will this source disappear due to the
higher order corrections? We discuss them from two aspects.

    (i) We have known that chaos in Eq. (1) origins from the
BFKL-singularity of the nonlinear evolution kernel. From the
experiences in the study of the BFKL equation, higher order QCD
corrections can not remove the singularities at the lower order
approximation. In general case, the infrared (IR) divergence in a
Feynman diagram at a given order can only be regularized by using
the virtual diagrams at the same order. We divide the higher order
corrections into two kinds: without and with the new IR-divergences.
We have examined in works [2,3] that the former can not remove the
chaos-GC solutions in Eq. (1), while the later may generate the
multi-chaos. Interestingly, these multi-chaos may combine into a
single chaos in the ZSR equation. Therefore, we believe that
chaos-GC effects still exist in Eq. (1) even considering the higher
order corrections

    (ii) We discuss the chaotic solution of Eq. (1) from the
view point of the chaos theory. It is well known that some of
chaotic attractors are unstable. A slight fluctuation of a parameter
may drive the system out of chaos. However, it has been proved that
some dynamical systems can exhibit robust chaos. A chaotic attractor
is said to be robust if, for its parameter values, there exist a
neighborhood in the parameter space with absence of periodic
negative Lyapunov exponents. Robustness implies that the chaotic
behavior cannot be destroyed by arbitrarily small perturbations of
the system parameters. The structure of the Lyapunov exponents shows
that chaos in Eq. (2.1) is robust [2]. The corrections from the
higher order processes cannot cancel the GC-effects.

    We presented the above two reasons to support the
GC-effects due to the special structure of the ZSR equation,
although we can't derive a complete modified ZSR equation, which
should include all order corrections. In this case, the evidence
from the experiments become important. Therefore, a following
question is where we can observed the GC-effects? We will discuss
the GC-effects at high energy cosmic rays in the following sections.

\newpage
\begin{center}
\section{The characters of the gluon condensation in cosmic ray spectra}
\end{center}

    The secondary particles in cosmic rays may origin from the hadronic precesses,
for example, $p+p(A)\rightarrow \pi+others$. The GC increases
suddenly the proton-proton or proton-nuclei cross sections. The
shape of energy spectra of these secondary particles is an ideal
subject to observe the GC-effects if the energy of incident proton
is accelerated beyond the GC-threshold $E_{p-p(A)}>E_{p-p(A)}^{GC}$.

    The cross section of inclusive gluon mini-jet production at high
energy $p-p$ collisions [20,22] reads

$$\frac{d\sigma_g}{d\underline{k}^2dy}=\frac{64N_c}{(N^2_c-1)\underline{k}^2}\int \underline{q} d
\underline{q}\int_0^{2\pi}
d\phi\alpha_s(\Omega)\frac{F(x_1,\frac{1}{4}(\underline{k}+\underline{q})^2)F(x_2,\frac{1}{4}
(\underline{k}-\underline{q})^2)}{(\underline{k}+\underline{q})^2(\underline{k}-\underline{q})^2},
\eqno(2)$$where
$\Omega=Max\{\underline{k}^2,(\underline{k}+\underline{q})^2/4,
(\underline{k}-\underline{q})^2/4\}$; The longitudinal momentum
fractions of interacting gluons are fixed by kinematics:
$x_{1,2}=\underline{k}e^{\pm y}/\sqrt{s}$.

    The number of secondary particles is proportional to Eq. (2)
if we neglect the mechanism of fragmentation of gluons into
secondary particles. We denote that $N_{\pi}(E_{p-p},E_{\pi})$ as
pion number with energies $E_{\pi}$ at $p-p$ collisions; $E_{p-p}$
is the energy of incident proton in the rest frame of targeted
proton. The calculations of the pion-distributions at the $p-p(A)$
collisions are very complicated due to the nonperturbative
hadroniztion. For the simplification, we consider only pions as the
secondary particles since the multiplicities of other particles at
high energy collisions are much smaller than that of pions. Usually,
these pions have smaller kinetic energy (or lower momentum) at the
center-of-mass (C.M.) system and form the central region in the
rapidity distribution. The maximum number of pions  $N_{\pi}$ at a
given interaction energy corresponds to a case, that all available
kinetic energies of pions at the C.M. system are almost used to
create pions. It leads to $N_{\pi}\sim \sqrt {s}$. However, the data
show that $N_{\pi}\sim \ln s$ or $\ln s^2$ [36]. A possible reason
is that the limited available number of gluons restricts the
increase of secondaries [37]. We assume that a large number of
gluons at the central region due to the GC-effects create the
maximum number $N_{\pi}$ of pions. We emphasize that this assumption
is a simplification method rather than a necessary GC-condition. In
fact, we will show that it may simplify the calculation but does not
essentially change the GC-characteristic signature. Using general
relativistic invariant and energy conservation, we have

$$(2m_p^2+2E_{p-p(A)}m_p)^{1/2}=E^*_{p1}+E^*_{p2}+N_{\pi}m_{\pi}, \eqno(3)$$
$$E_{p-p(A)}+m_p=m_p\gamma_1+m_p\gamma_2+N_{\pi}m_{\pi}\gamma, \eqno(4)$$where
$E^*_{p_i}$ is the energy of leading proton at the C.M. system,
$\gamma_i$ are the corresponding Lorentz factors. Using the
inelasticity $K\sim 0.5$ [38], we set

$$E^*_{p1}+E^*_{p2}=(\frac{1}{K}-1)N_{\pi}m_{\pi},  \eqno(5)$$and

$$m_p\gamma_1+m_p\gamma_2=(\frac{1}{K}-1) N_{\pi}m_{\pi}\gamma. \eqno(6)$$
One can easily get the solutions $N_{\pi}(E_{p-p(A)},E_{\pi})$ for
the $p-p(A)$ collisions

$$\ln N_{\pi}=0.5\ln E_{p-p(A)}+a, ~~\ln N_{\pi}=\ln E_{\pi}+b,~~ where~E_{\pi}
\in [E_{\pi}^{GC},E_{\pi}^{max}]. \eqno(7)$$ The parameters

$$a\equiv 0.5\ln (2m_p)-\ln m_{\pi}+\ln K, \eqno(8)$$ and

$$b\equiv \ln (2m_p)-2\ln m_{\pi}+\ln K. \eqno(9)$$Equation (7) gives
the relation among $N_{\pi}$, $E_{p-p(A)}$ and $E_{\pi}^{GC}$ by
one-to-one, which leads to a GC-characteristic spectrum.

\newpage
\begin{center}
\section{The GC effects in the gamma ray spectra}
\end{center}

 Imaging a high energy proton collides with a proton or a
nucleus, we have $p+p(A)\rightarrow \pi^{\pm,0}+others$ and
following $\pi^0\rightarrow 2\gamma$. The corresponding gamma flux
in a general hadronic framework reads

$$\Phi_{\gamma}(E_{\gamma})=\Phi^0_{\gamma}(E_{\gamma})+
\Phi^{GC}_{\gamma}(E_{\gamma}),\eqno(10)$$$\Phi^0_{\gamma}(E_{\gamma})$
is the background contribution and

$$\Phi^{GC}_{\gamma}(E_{\gamma})=C_{p-p(A)}\left(\frac{E_{\gamma}}{E_0}\right)^{-\beta_{\gamma}}
\int_{E_{\pi}^{min}}^{E_{\pi}^{max}}dE_{\pi}
\left(\frac{E_{p-p(A)}}{E_{p-p(A)}^{GC}}\right)^{-\beta_p}N_{\pi}(E_{p-p(A)},E_{\pi})
\frac{d\omega_{\pi-\gamma}(E_{\pi},E_{\gamma})}{dE_{\gamma}},
\eqno(11)$$ where index $\beta_{\gamma}$ and $\beta_p$ denote the
propagating loss of gamma rays and the acceleration mechanism of
protons respectively; $C_{p-p(A)}$ incorporates the kinematic factor
with the flux dimension and the percentage of $\pi^0\rightarrow
2\gamma$. The normalized spectrum for $\pi^0\rightarrow 2\gamma$ is

$$\frac{d\omega_{\pi-\gamma}(E_{\pi},E_{\gamma})}{dE_{\gamma}}
=\frac{2}{\beta_{\pi}
E_{\pi}}H[E_{\gamma};\frac{1}{2}E_{\pi}(1-\beta_{\pi}),
\frac{1}{2}E_{\pi}(1+\beta_{\pi})], \eqno(12)$$ $H(x;a,b)=1$ if
$a\leq x\leq b$, and $H(x;a,b)=0$ otherwise. Inserting Eq. (7) and
(12) into Eq. (11), we have

$$\Phi^{GC}_{\gamma}(E_{\gamma})=C_{p-p(A)}\left(\frac{E_{\gamma}}{E_{\pi}^{GC}}\right)^{-\beta_{\gamma}}
\int_{E_{\pi}^{GC}~or~E_{\gamma}}^{E_{\pi}^{GC,max}}dE_{\pi}
\left(\frac{E_{p-p(A)}}{E_{p-p(A)}^{GC}}\right)^{-\beta_p}N_{\pi}(E_{p-p(A)},E_{\pi})
\frac{2}{\beta_{\pi}E_{\pi}}, \eqno(13)$$where the lower-limit of
the integration takes $E_{\pi}^{GC}$ (or $E_{\gamma}$) if
$E_{\gamma}\leq E_{\pi}^{GC}$ (or if $E_{\gamma}> E_{\pi}^{GC}$). In
consequence,

$$E_{\gamma}^2\Phi^{GC}_{\gamma}(E_{\gamma})=\left\{
\begin{array}{ll}
\frac{50C}{2\beta_p-1}(E_{\pi}^{GC})^3\left(\frac{E_{\gamma}}{E_{\pi}^{GC}}\right)^{-\beta_{\gamma}+2} & {\rm if~}E_{\gamma}\leq E_{\pi}^{GC}\\\\
\frac{50C}{2\beta_p-1}(E_{\pi}^{GC})^3\left(\frac{E_{\gamma}}{E_{\pi}^{GC}}\right)^{-\beta_{\gamma}-2\beta_p+3}
& {\rm if~} E_{\gamma}>E_{\pi}^{GC}
\end{array} \right. .\eqno(14)$$It is the power-law with a sharp break.
The break energy $E_{\gamma}=E_{\pi}^{GC}$ is a direct result of the
gluon distribution in Fig. 1, where a sharp peak divides the
spectrum into two parts.

    A pure power-law at $E_{\gamma}<E_{\pi}^{GC}$ origins from a fixed
lower limit $\sim E_{\pi}^{GC}$ of integral in Eq. (13) and it is
irrelevant to the concrete form of Eq. (7). The above two universal
behaviors of $\Phi_{\gamma}^{GC}$ are directly arisen from the
GC-effects and they are different from all other well known smoothly
radiation spectra. We regard them as the GC-character. The second
power-law at $E_{\gamma}>E_{\pi}^{GC}$ is a simplified result in
Eqs. (3) and (4), where all available kinetic energies in the
central region are used to create pions. We emphasize that any
deformations from this power-law at $E_{\gamma}>E_{\pi}^{GC}$ are
allowed if our simplification is modified. One can compare Eq. (14)
with the experiments to check the validity of the simplification.
Even so, it does not change the above mentioned GC-character.

    The SNRs are important gamma ray sources. There are different shapes
of gamma energy spectra, which corresponds to various production
mechanisms. For example, gamma rays can be generated as
bremsstrahlung radiation when electrons and positrons interact with
ambient matter, or as a result of inverse Compton scattering of low
energy photons. Hadronic interactions can also create gamma rays.
Indeed, proton-proton (or proton-nuclei and nuclei-nuclei)
collisions may create $\pi^0$ meson, which quickly decays into two
gamma photons. The recent detection of the neutral pion-decay
signature from two middle-aged SNRs: IC 443 and W44 has been
demonstrated [39,40]. In normal case the gamma-ray spectrum
$\Phi_{\gamma}(E_{\gamma})$ is symmetric about
$E_{\gamma}=m_{\pi}^0/2=67.5$ MeV. However, we will show that the
GC-effects break the gamma power-law at GeV-TeV band. and induce the
GC-effects in their gamma spectra.

     Tycho is located in a relatively clean environment and have been
studied over a wide range of energies. Cosmic ray protons
accelerated in SNRs may reach the $E_{p-p(A)}^{GC}$-energy range and
they interact with the surrounding matter inside SNR. Thus, the
GC-effects appear in their energy spectra. Figure 5 gives our
predicted gamma spectrum and the comparison with Tycho's spectrum
[4], where we take two GC-sources $E_{\pi}^{GC}=24$ TeV for $p-p$
and $E_{\pi}^{GC}=800$ GeV for $p-A$. Note that $E_{\pi}^{GC}=24$
TeV corresponds to $E_{p-p}^{GC}=6\times 10^{10}$ GeV$=E_{GZK}$ or
$\sqrt{s_{p-p}^{GC}}=3.3\times 10^5$ GeV. One can find that a
power-law at $E_{\gamma}<800$ GeV and its sharp break due to the
GC-effects are significant distinction from other models. Note that
we use Eq. (14) at $E_{\gamma} > 100$ MeV where the energy loss of
photons mainly origin from pairproduction. We need to consider
Compton scattering and photoelectron effect at more lower energy
range and change the index $\beta_{\gamma}$.

   \begin{figure}
   	\begin{center}
   		\includegraphics[width=0.8\textwidth]{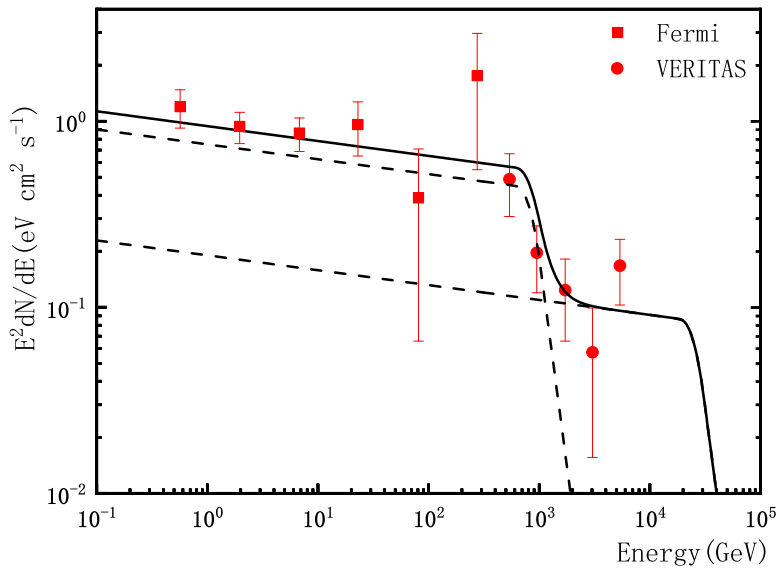} 
   		\caption{Predicted gamma ray spectra multiplied by $E^2$
   			and comparisons with the VERITAS- and Fermi-spectrum for Tycho's
   			supernova remnant [4]. The broken lines are the p-A and p-p
   			contributions, the solid line is the total contributions of the
   			GC-effects. The parameters see Table 1.}\label{fig:5}
   	\end{center}
   \end{figure}

    The sharp broken power-law was observed by Fermi-LAT and H.E.S.S. in
several AGNs and SNRs. This phenomenon was explained as the emitted
gamma-rays suffer from absorption by the extragalactic background
light (EBL) during their travel towards Earth from the sources.
However, both the distribution of EBL photons and the intrinsic VHE
spectra of distant sources are unknown. The GC-effects provide an
alternative possible explanation, where the intrinsic broken
spectrum is fixed and the gamma propagating corrections including
EBL absorption are described by power-law $\sim
E_{\gamma}^{-\beta_{\gamma}}$ at $E_{\gamma}>100$ GeV.

     H.E.S.S. observed the high-frequency peaked of brizars PKS
2155-304 and PG 1553-113, which are among the brightest objects in
the VHE gamma-ray sky [41]. Combining the spectra derived from
Fermi-LAT data, the results indicate a sharp break in the observed
spectra of both sources at $E_{\gamma}\sim 100$ GeV. Our predicted
spectra and comparisons with the data are shown in Figs.6 and 7. The
parameters are listed in Table 1. The GeV gamma-ray spectrum of
supernova remnants H.E.S.S. J1731-347 and SN 1006 combining
Fermi-LAT data show a broken power-law once again at $E_{\gamma}\sim
1$ TeV [42]. We fit it using the Eq. (14) in Fig. 8 and 9.  Figures
10 and 11 are the spectra of PKS 2005-489 and Mrk 421 [43] and the
comparisons with Eq. (14).

\begin{figure}
	\begin{center}
		\includegraphics[width=0.8\textwidth]{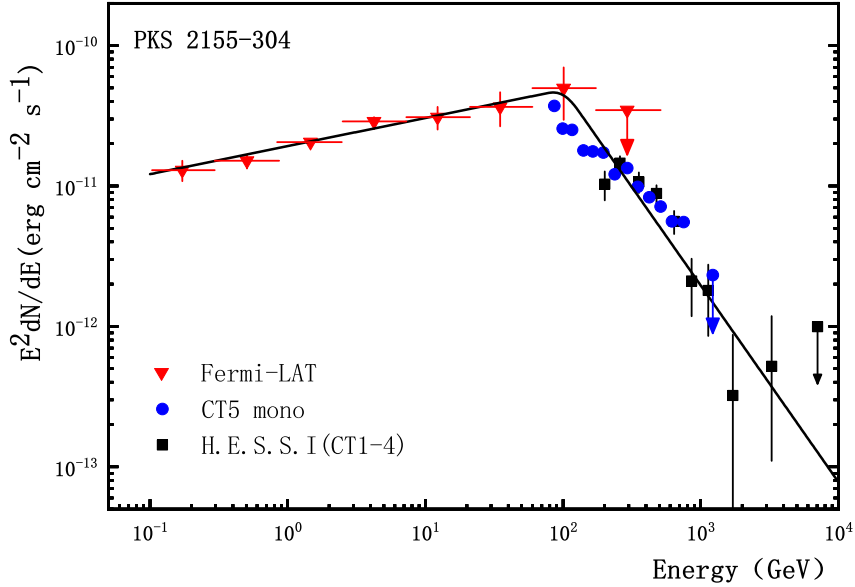} 
		\caption{ Predicted gamma ray spectrum multiplied by $E^2$
			and comparisons with the H.E.S.S. and Fermi-LAT spectrum for PKS
			2155-304 [40]. The parameters see Table 1.}\label{fig:6}
	\end{center}
\end{figure}

\begin{figure}
	\begin{center}
		\includegraphics[width=0.8\textwidth]{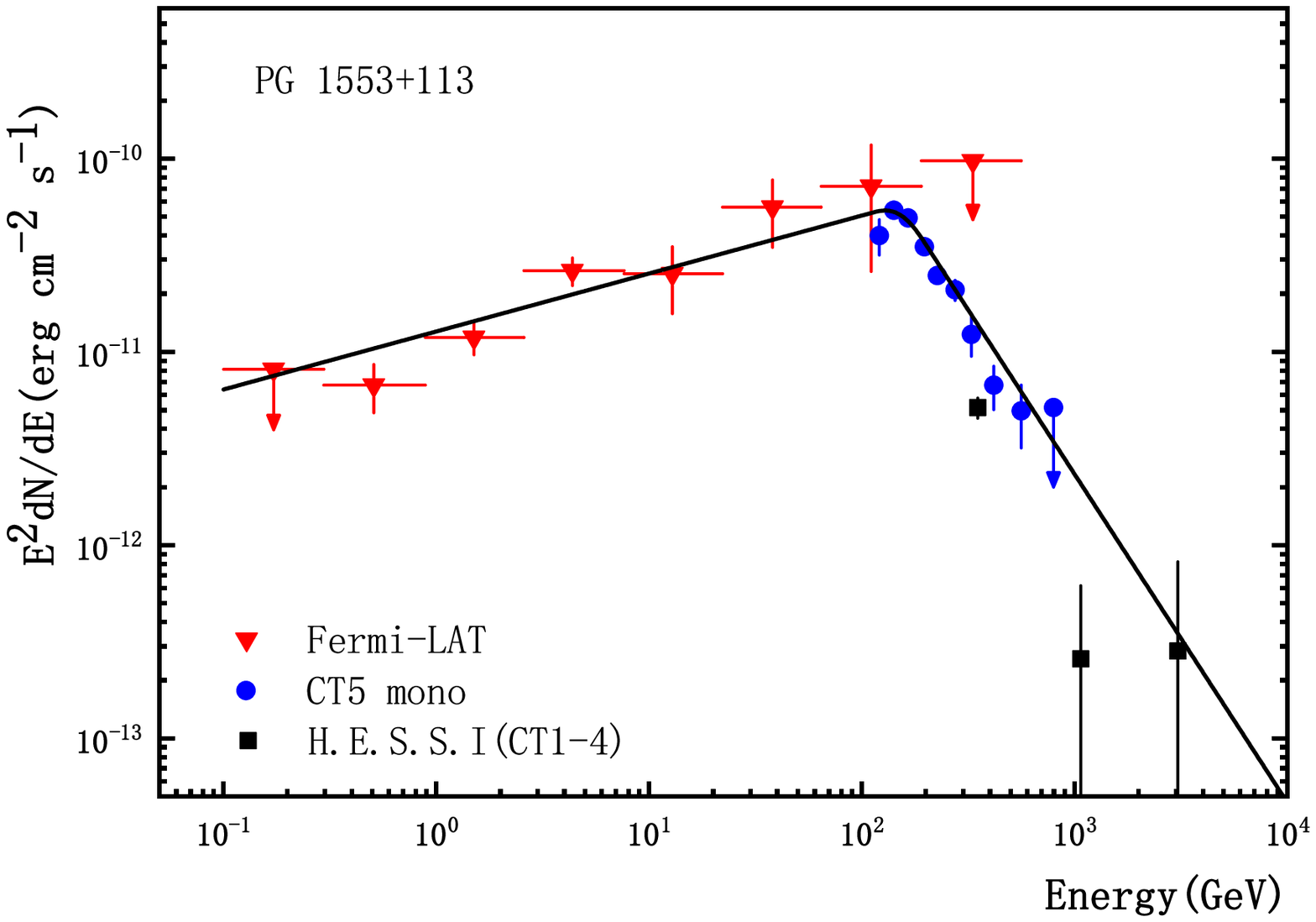} 
		\caption{Predicted gamma ray spectrum multiplied by $E^2$
			and comparisons with the H.E.S.S. and Fermi-LAT spectrum for PG
			1553+113 [40]. The parameters see Table 1.}\label{fig:7}
	\end{center}
\end{figure}

\noindent \centerline{Table 1: The parameters of gamma-ray spectra.}

\includegraphics[width=0.9\textwidth]{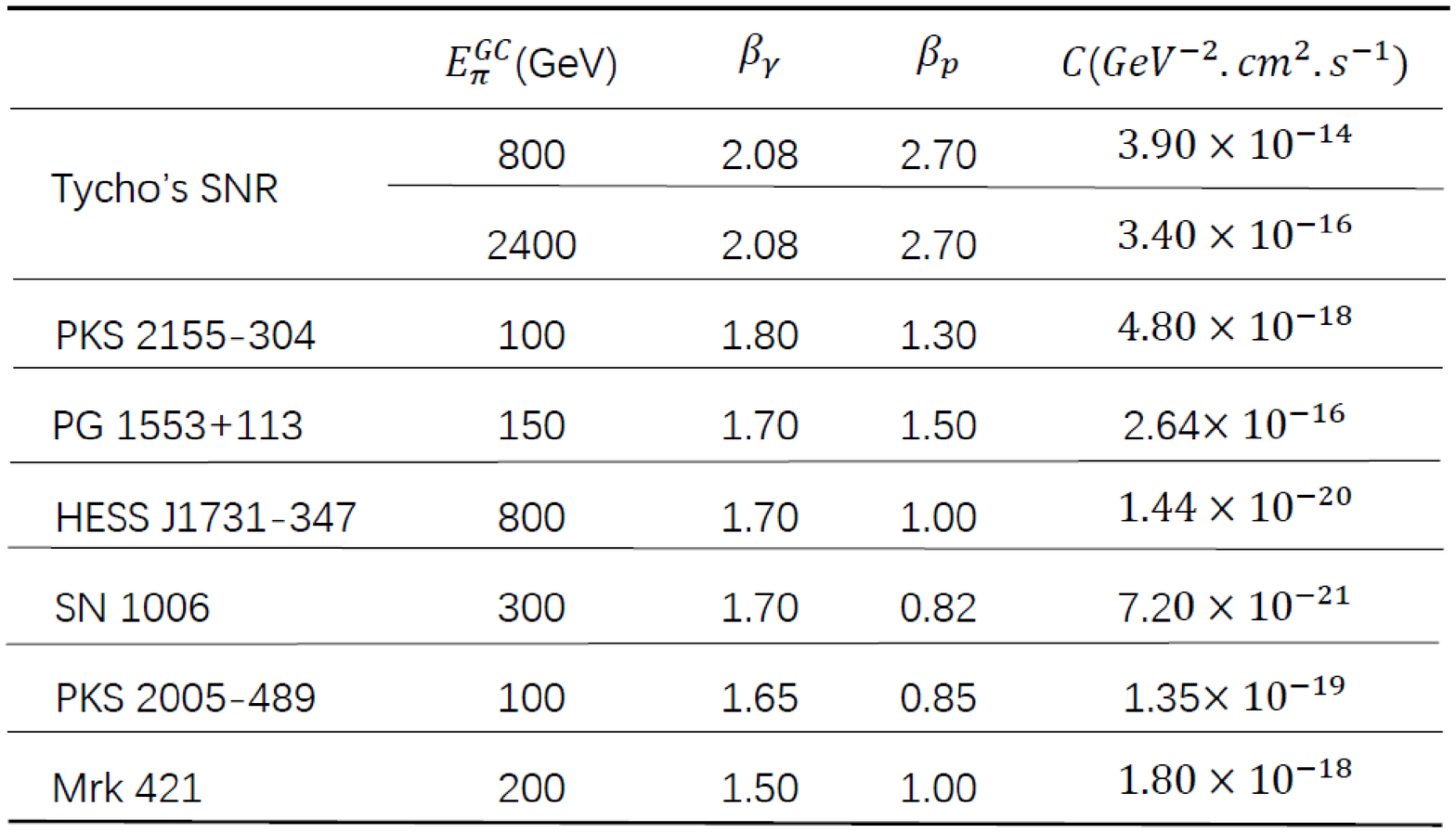} 

     We discuss the parameters in the above examples.
Usually, the power indexes of cosmic rays spectra contains the
contributions of the production mechanism and the propagation effect
and they take as  $\beta_{\gamma}\sim 1.5-2$ and $\beta_p\sim 2-3$.
Note that these values of $\beta_p$ contains the interstellar
propagation corrections of proton, while in this work primary
protons are restricted inside the source. Therefore, the above
smaller indexes are accepted.

    Now we turn to the GC-threshold energy. Greisen-Zatsepin-Kuzmin
(GZK) [44,45] predicted a drastic reduction of the spectrum of
cosmic proton rays near the energy $E_{GZK}\sim 6\times 10^{19}$ eV,
since the energy of the cosmic rays losses in the collisions with
cosmic microwave background radiation during their long propagation.
Considering that $E_{GZK}$ is a characteristic scale, we expect that
the GC- effects begin from the incident proton energy
$E_{p-p}^{GC}\equiv E_{GZK}=6\times 10^{19}$ eV, or
$\sqrt{s_{p-p}^{GC}}=330$ TeV. We emphasize that this is only a
possible choice and $E_{p-p}^{GC}>E_{GZK}$ is also possible.
However, it does not affect the following discussion, since there
are only a few available data with large uncertainty at this energy
range.
\begin{figure}
	\begin{center}
		\includegraphics[width=0.8\textwidth]{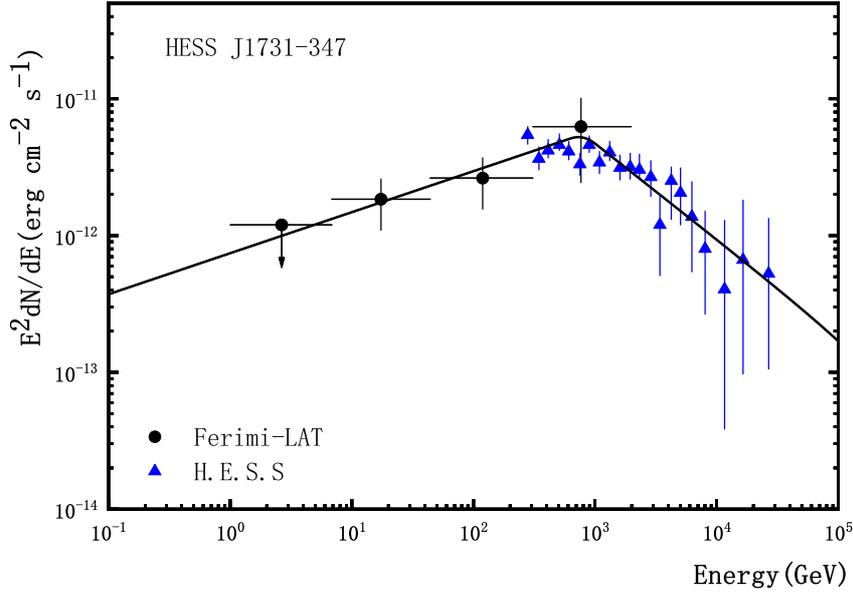} 
		\caption{Predicted gamma ray spectra multiplied by $E^2$
			and comparisons with the H.E.S.S. and Fermi-LAT for J1731 [41]. The
			parameters see Table 1.}\label{fig:8}
	\end{center}
\end{figure}

\begin{figure}
	\begin{center}
		\includegraphics[width=0.8\textwidth]{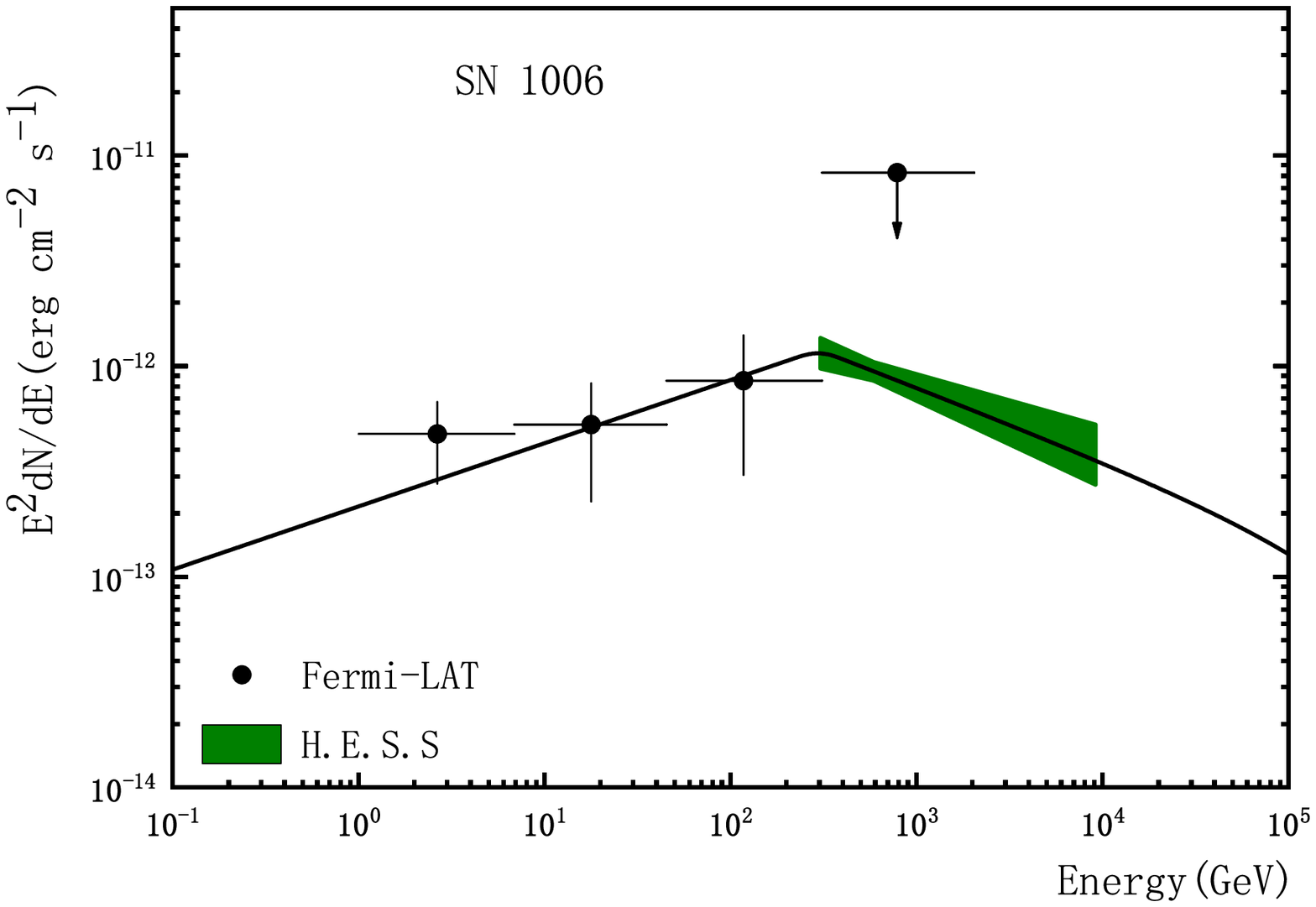} 
		\caption{Predicted gamma ray spectrum multiplied by $E^2$
			and comparisons with the H.E.S.S. and Fermi-LAT spectrum for SN 1006
			[41]. The parameters see Table 1.}\label{fig:9}
	\end{center}
\end{figure}

\begin{figure}
	\begin{center}
		\includegraphics[width=0.8\textwidth]{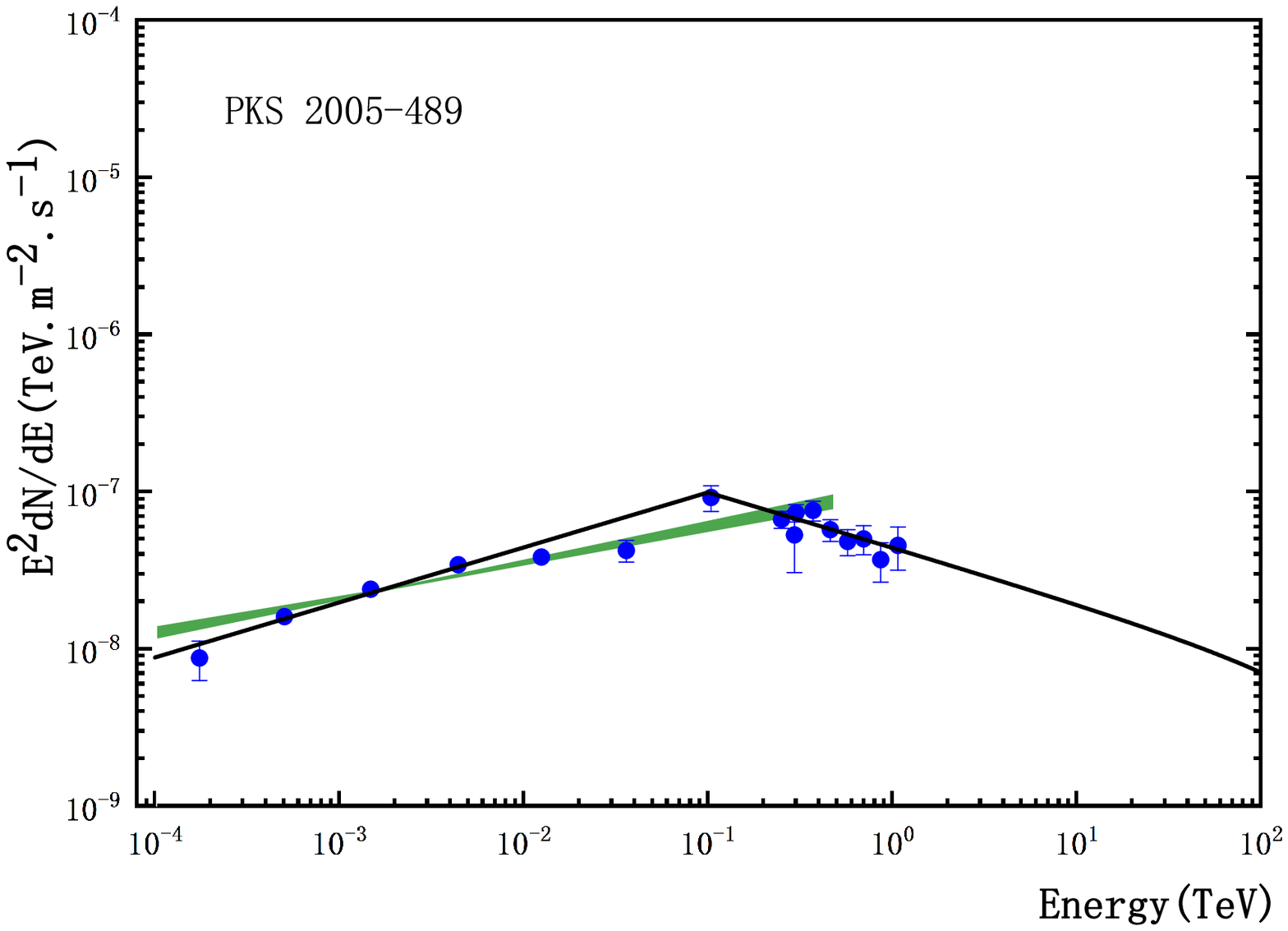} 
		\caption{Predicted gamma ray spectrum multiplied by
			$E^2$ and comparisons with the H.E.S.S. and Fermi-LAT spectrum for
			PKS 2005 [42]. The parameters see Table 1.}\label{fig:10}
	\end{center}
\end{figure}

\begin{figure}
	\begin{center}
		\includegraphics[width=0.8\textwidth]{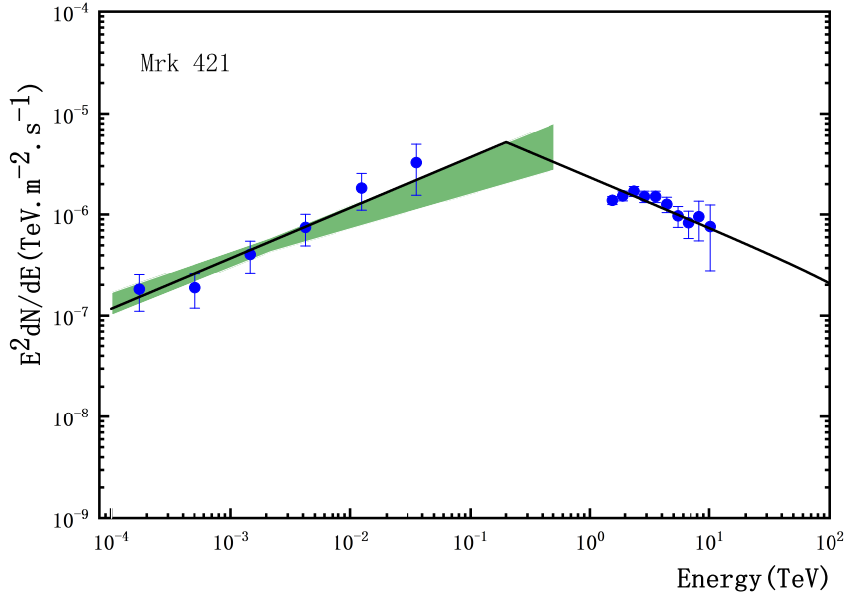} 
		\caption{Predicted gamma ray spectrum multiplied by
			$E^2$ and comparisons with the H.E.S.S. and Fermi-LAT spectrum for
			Mrk 421 [42]. The parameters see Table 1.}\label{fig:11}
	\end{center}
\end{figure}

    The cross section $\sigma_{p-air}$ was derived using the
distribution of the shower maximum slant depth $X_{max}$, where the
interaction length and consequently $\sigma_{p-air}$ relate to the
$X_{max}$ distribution's exponential tail in the cascade model. The
result of Telescope Array has not found a big increment effect in
$\sigma_{p-air}$ till $\sqrt{s}=95$ TeV [46,47]. However, the
cascade method has the fundamental limitation. Some of important
parameters are undetermined [48]. Besides, the GC-effects may disappear
quickly at the beginning collisions due to the energy loss of the
leading proton. Therefore, we do not regard the Telescope Array data
as a restriction to the value of $E_{\pi}^{GC}$ in this work

     The GC-threshold $E_{p-p(A)}^{GC}$ is target-dependent.
Because the nonlinear term of the QCD evolution equation should be
re-scaled by $A^{1/3}$, $E_{\pi}^{GC}$ decreases with increasing A
[2]. High energy protons may collide with nuclei in Universe. The
GC-threshold decreases in high dense (or heavy nuclear) matter
[2,3]. Therefore, $E_{p-A}^{GC}<E_{p-p}^{GC}$ is expected.
A-dependence of $E_{p-A}^{GC}$ is a complicated problem, which
relates to the distribution and structure of the GC-source. In this
work we use the experimental data to determine the values of
$E_{p-p(A)}^{GC}$ (or $E_{\pi}^{GC}$). However, we should point out
that (i) not all elements show their GC-effects, some of them will
be shadowed by the background due to lower abundance or suppression
factors in the propagation.

\newpage
\begin{center}
\section{The GC-effects in the electron-positron spectra}
\end{center}

    High energy photons via $p+p(A)\rightarrow \pi$ and $\pi^0\rightarrow \gamma$
through strong electric field inside source, they may product
electron/ positron pair.

    Considering the protons are accelerated to beyond the GC-threshold
inside a strong source (for say, supernova remnant). The primary
product of hadronic processes is the production of pion via

$\pi=(\pi^+,\pi^0,\pi^-)$. Then we have $\pi^{\pm}\rightarrow
\mu^{\pm}+\nu_{\mu}(\overline{\nu}_{\mu})$, $\mu^{\pm}\rightarrow
e^{\pm}+\nu_e(\overline{\nu}_e)+\overline{\nu}_{\mu}({\nu}_{\mu})$
and $\pi^0\rightarrow 2\gamma$. We focus the elelctron/positron pair
cascades $\gamma + magnetic field\rightarrow e^++e^-$ in the source.
We use the hadronic model to calculate the electron/positron
spectra, which is combined by includes the following factors: the
spectrum of injection proton $E_{p-p(A)}^{-\beta_p}$, $\pi$-spectrum
$N_{\pi}$ at $p-p(A)$ collisions beyond the GC-threshold, the
probabilities of $\pi^0\rightarrow \gamma$ and $\gamma\rightarrow
e^-+e^+$, the electron/posotron propagation effects
$E_j^{-\beta_j}$. The isotropic measured electron and positron
fluxes are

$$\Phi_j(E_j)=\Phi^0_j(E_j)+\Phi^{GC}_j(E_j),\eqno(15)$$ for $j=e^-$ or $e^+$
and
$$\Phi_j^{GC}(E_j)=C_j\left(\frac{E_j}{E_0}\right)^{-\beta_j}
\int_{E_j}dE_{\gamma}\left(\frac{E_{\gamma}}{E_0}\right)^{-\beta_{\gamma}}
\int_{E_{\pi}^{min}}^{E_{\pi}^{max}}
dE_{\pi}\left(\frac{E_{p-p(A)}}{E_{p-p(A)}^{GC}}\right)^{-\beta_p}$$
$$N_{\pi}(E_{p-p(A)},E_{\pi})\frac{d\omega_{\pi-\gamma}(E_{\pi},E_{\gamma})}{dE_{\gamma}}
\frac{d\omega_{\gamma-e}(E_{\gamma}, E_e)}{dE_e}$$
$$=C_j\left(\frac{E_j}{E_{\pi}^{GC}}\right)^{-\beta_j}\int_{E_j}\frac{dE_{\gamma}}{E_{\gamma}}
\left(\frac{E_{\gamma}}{E_{\pi}^{GC}}\right)^{-\beta_{\gamma}}
\int_{E_{\pi}^{GC}~or~E_{\gamma}}^{E_{\pi}^{max}}
dE_{\pi}\left(\frac{E_{p-p(A)}}{E_{p-p(A)}^{GC}}\right)^{-\beta_p}N_{\pi}(E_{p-p(A)},E_{\pi})
\frac{2}{\beta_{\pi}E_{\pi}}$$
$$=\left\{
\begin{array}{ll}
\frac{50C_j}{2\beta_p-1}E_{\pi}^{GC}\left(\frac{E_j}{E_{\pi}^{GC}}\right)^{-\beta_j}\left[\frac{1}{\beta_{\gamma}}
\left(\frac{E_j}{E_{\pi}^{GC}}\right)^{-\beta_{\gamma}}+
(\frac{1}{\beta_{\gamma}+2\beta_p-1}-\frac{1}{\beta_{\gamma}})\right]
& {\rm if~} E_j\leq E_{\pi}^{GC}\\\\
\frac{50C_j}{(2\beta_p-1)(\beta_{\gamma}+2\beta_p-1)}
(E_{\pi}^{GC})\left(\frac{
E_j}{E_{\pi}^{GC}}\right)^{-\beta_j-\beta_{\gamma}-2\beta_p+1} &
{\rm if~} E_j>E_{\pi}^{GC}
\end{array} \right. \eqno(16)$$where the integral lower-limit takes $E_{\pi}^{GC}$ (or
$E_{\gamma}$) if $E_{\gamma}\leq E_{\pi}^{GC}$ (or if $E_{\gamma}>
E_{\pi}^{GC}$). After taking average over possible directions, the
energy of pair-produced electron-positron is uniformly distributed
from zero to maximum value, i.e.,
$$\frac{d\omega_{\gamma-e}(E_{\gamma},
E_e)}{dE_e}=\frac{1}{E_{\gamma}}.\eqno(17)$$ In Eq. (16)
$N_{\pi}(E_{p-p(A)},E_{\pi})$ is calculated using Eq. (5);
$E_{p-p(A)}$ is the energy of incident proton in the rest frame of
targeted proton. Equation (16) shows the electron spectrum is
smoothly broken at $\sim E_{\pi}^{GC}$.

\begin{figure}
	\begin{center}
		\includegraphics[width=0.8\textwidth]{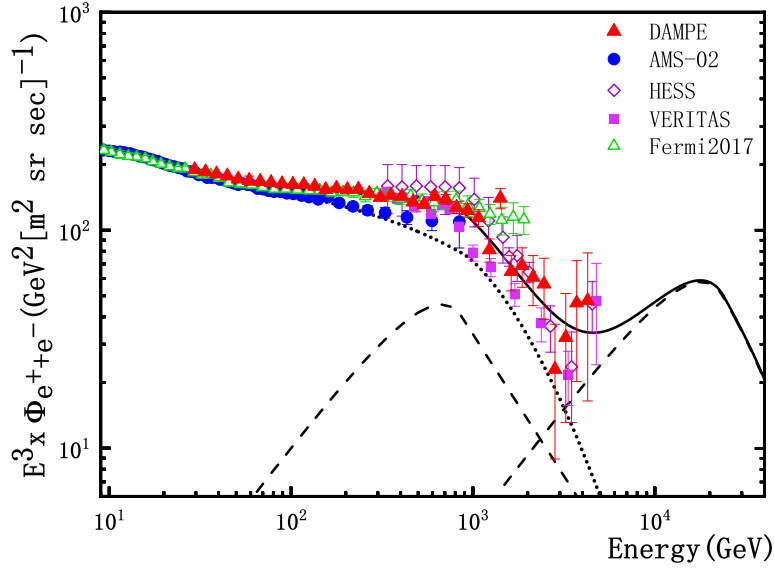} 
		\caption{Predicted cosmic electron spectrum multiplied
			by $E^3$ as a function of energy (solid curve). Broken curves
			present the broken power-law of the GC-effects. The data are taken
			from [1,4-7].}\label{fig:12}
	\end{center}
\end{figure}

   Figure 12 is a result fitting data including the DAMPE spectrum,
where using $E_{\pi}^{GC}=880$ GeV for $p-A$ collisions and
$E_{\pi}^{GC}=24$ TeV for $p-p$ collisions, respectively.
$\Phi_{e^-+e^+}^0$ is refers to [49], where we reduce the background
line to lower than the data at $E>1$ TeV. The data are taken from
[49-52]. The parameters $E_{\pi}^{GC}=880$ GeV, $\beta_p=1.7,
\beta_\gamma=1.3, \beta_e=0.6, C_{880~GeV}=1.15\times10^{-6}$ and
$C_{24~TeV}=3.6\times10^{-9}$. The results present a smoothly broken
power at 0.9 TeV and the curve is turning again at $3\sim 4$ TeV.
Usually, the power index $\beta_e\sim 2-3$. The production mechanism
of electron/posotron is separated by $d\omega_{\gamma-e}/{dE_e}$.
Therefore, the above smaller index is accepted.

    For understanding the GC-effects in this example, we image the case without the GC-effects.
The gamma-ray spectrum from $p+p\rightarrow \pi^0\rightarrow
2\gamma$ will peak at $\sim 1$ GeV and it has been proved [33,34].
In this case, the following contributions of $\gamma\rightarrow
e^++e^-$ at tail ($>100$ GeV) to the electron-flux are much lower
than the background and they are completely negligible.

    Why the GC-threshold $E_{p-A}^{GC}$ (or $E_{\pi}^{GC}$) of gamma-ray
spectra is more abundant than that of electron-positron spectra. The
reason is that the observed charged cosmic rays on the Earth origin
mainly from the nearest galaxy-milky way. Most electron/positron
origin from extragalactic galaxies (AGNs) and they are suppressed
due to radiation in long distance transmission.

    In concretely, we assume that the GC-threshold at $p-Pb$ collisions
is $E_{p-Pb}^{GC}=3.4\times 10^7$ GeV or $\sqrt{s_{p-Pb}^{GC}}=
\sqrt{s_{Pb-Pb}^{GC}}=8$ TeV. Thus, we suggest that
$E_{\pi}^{GC}\simeq 1$ TeV in electron/postron spectra origins from
proton-intermediate nucleus. While the GC-threshold $E_{\pi}^{GC}<<
1$ TeV in gamma ray spectra may arise from proton-heavy nucleus
collision or proton-(head-on) moving nucleus. In the later case, the
effective GC-threshold is reduced.

    A large amount of pions with a certain energy accumulate in a narrow
space at per collision, they may transform each other in the
formation time due to their wave-functions overlap, i.e.,
$\pi^++\pi^-\rightleftharpoons 2\pi^0$. However, the above balance
will be broken since $m_{\pi^+}+m_{\pi^-}>2m_{\pi^0}$, and the
lifetime of $\pi^0$ ($10^{-16}$ s) is much shorter than the typical
weak decay lifetimes of $\pi^{\pm}$ ($10^{-6}~s-10^{-8}~s$).
Therefore we neglect temporarily the contributions of $\pi^{\pm}$.
The valid of this assumption will be checked by the following
discussions

    Now we consider the contributions of
$\pi^{\pm}\rightarrow \mu^{\pm}+\nu_{\mu}(\overline{\nu}_{\mu})$ and
$\mu^{\pm}\rightarrow
e^{\pm}+\nu_e(\overline{\nu}_e)+\overline{\nu}_{\mu}({\nu}_{\mu})$,
although this process is negligible as discussed above. Similar to
Eq. (16), we have

    $$\Phi_e^{GC}(E_e)=C_e\left(\frac{E_e}{1~GeV}\right)^{-\beta_e}
\int
dE_{\mu}\int_{2.5E_e~or~E_{\pi}^{GC}}^{E_{\pi}^{max}}\frac{dE_{\pi}}{E_{\pi}}
\left(\frac{E_{p-p(A)}}{E_{p-p(A)}^{GC}}\right)^{-\beta_p}N_{\pi^{\pm}}(E_{p-p(A)},E_{\pi})$$
$$\frac{d\omega_{\pi-\mu}(E_{\pi},E_{\mu})}{dE_{\mu}}\frac{d\omega_{\mu-e}(E_{\mu},E_e)}{dE_e}$$
$$=\left\{
\begin{array}{ll}
\frac{370C_e}{2\beta_p-1}E_e^{GC}\left(\frac{E_e}{E_e^{GC}}\right)^{-\beta_e+2} & {\rm if~}E_e\leq E_e^{GC}\\\\
\frac{370C_e}{2\beta_p-1}E_e^{GC}\left(\frac{E_e}{E_{e}^{GC}}\right)^{-\beta_e-2\beta_p+2}
& {\rm if~} E_{\gamma}>E_e^{GC}
\end{array} \right., \eqno(18)$$where the
normalized spectra are
$$\frac{d\omega_{\pi-\mu}(E_{\pi},E_{\mu})}{dE_{\mu}}=\delta (E_{\mu}-0.8E_{\pi}),
\eqno(19)$$ and
$$\frac{d\omega_{\mu-e}(E_{\mu},E_e)}{dE_e}=4(\frac{2E_e}{E_{\mu}})^2(1.5-\frac{2E_e}{E_{\mu}}),~~E_e\leq
\frac{E_{\mu}}{2}. \eqno(20)$$

    The results are shown in Fig. 13, where we take
break energy $E_e^{GC}\sim 1.4$ TeV. The spectrum in Fig. 13
presents a sharp broken power-law, which is different from a
smoothed broken power at $E_{\pi}^{GC}\sim 0.9$ TeV. The reason is
that the integral of $E_{\gamma}$ in Eq. (16) smooths the corner,
while Eq. (14) lacks such smooth factor. .

   The angle at 1.4 TeV may become more smaller if we assume that
there is dense gas around this GC-source. In this case, a larger
diffusion index will be incorporated into the index $\beta_p$ in Eq.
(19) when the injected protons crossing the dense gas region. On the
other hand, a suppressed factor $B$ should be inserted into Eq.
(19), which describes that a part of lower energy electrons are
stopped in the dense gas. Figure 14 shows a result. This example
shows that a sharp peak at the TeV-band seems permissible. However,
as we have mentioned that the probability of $\pi^{\pm}$ decay is
much smaller than that of $\pi^0$ decay due to the accumulation of
pions at $p-p(A)$ collisions. Therefore, we suggest that the sharp
structure is appeared as a rare event and it maybe mixed with random
fluctuations in the measurements.

\begin{figure}
	\begin{center}
		\includegraphics[width=0.8\textwidth]{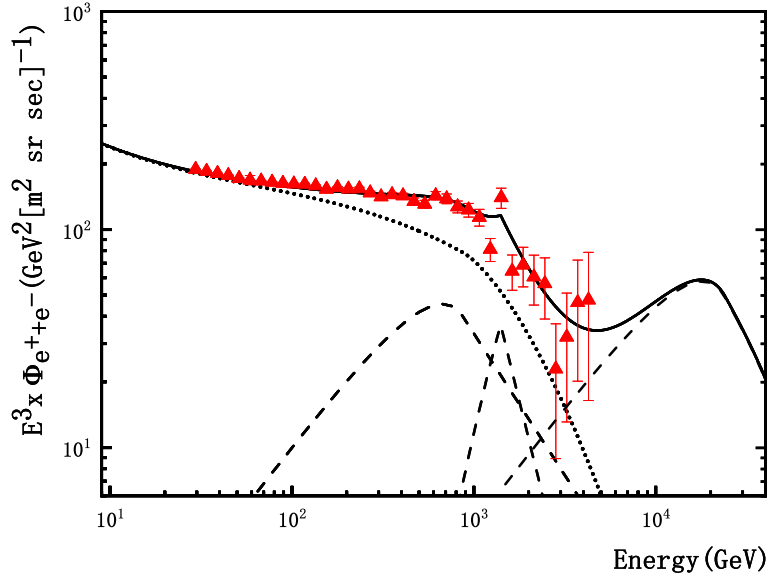} 
		\caption{As similar to Fig. 12 but added the corrections
			of Eq. (3.5). The added parameters for the peak at $1.4~TeV$:
			$\beta_{p}=3.9$, which incorporates the index of $\pi^{\pm}$,
			$\beta_e=0.6$ and $C_{1.4~TeV}=5.0\times10^{-15}$. The data are
			taken from the DAMPE [47].}\label{fig:13}
	\end{center}
\end{figure}

\begin{figure}
	\begin{center}
		\includegraphics[width=0.8\textwidth]{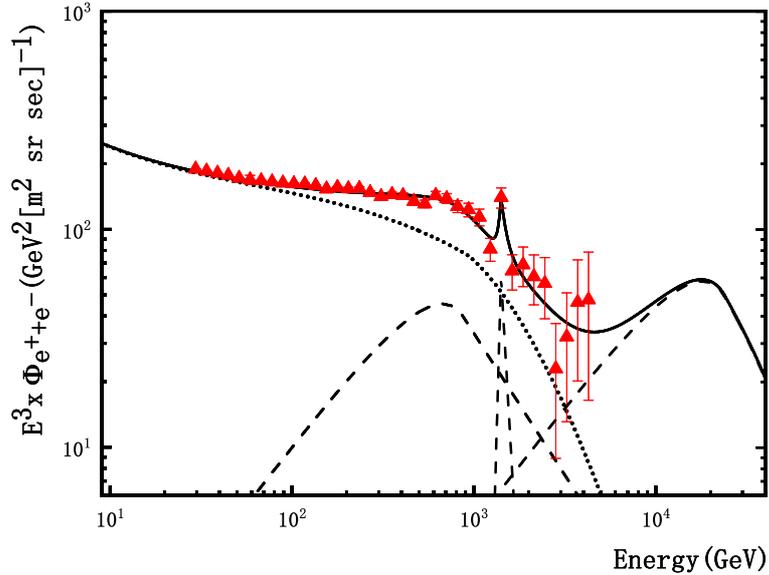} 
		\caption{Similar to Fig. 13 but added a suppression
			factor $B=\theta(1.4-E_e)\exp(-0.2\times(1.4-E_e))$ in GeV unit and
			a larger effective $\beta_p=3.9+5.6$ are used due to the corrections
			of the imaginary dense gas around this GC-source.}\label{fig:14}
	\end{center}
\end{figure}

\begin{figure}
	\begin{center}
		\includegraphics[width=0.8\textwidth]{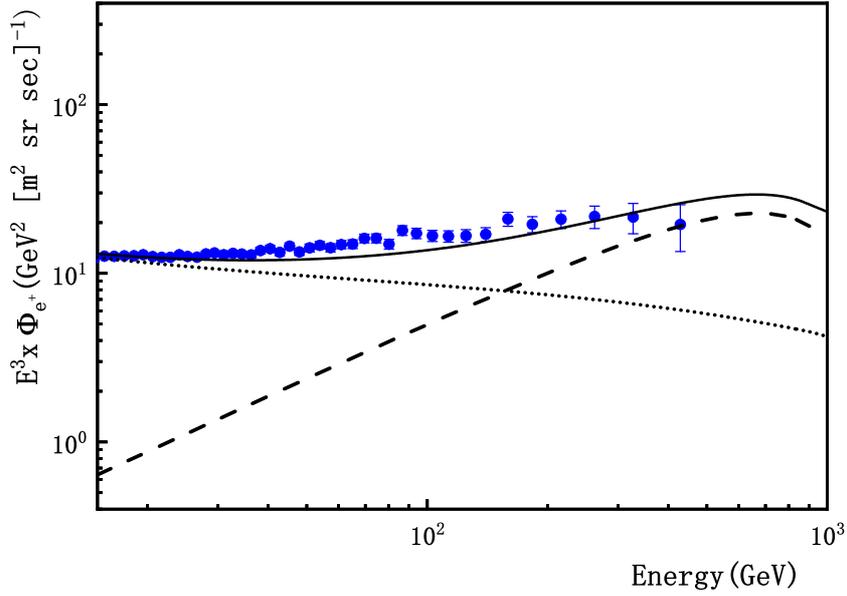} 
		\caption{Predicted positron spectrum multiplied by $E^3$
			as a function of energy. The data are taken from [53-56].}\label{fig:15}
	\end{center}
\end{figure}

    The positron spectrum could show GC-signature clearer because a very low $e^+$-background.
The electron and positron have an equal creation probability from
$\gamma\rightarrow e^++e^-$, i.e.,

    $$\Phi_{e^+}^{GC}(E_e)=\frac{1}{2}\Phi_{e^-+e^+}^{GC}(E_e).\eqno(21)$$
However, the background of positron flux is much lower than that of
electron flux in sky. It origins from the unknown cosmic evolution.
Unfortunately, we lack the data of positron background, which is
model-dependent. We take a simple assumption, i.e.,

    $$\Phi_{e^+}^0(E_e)=\eta\Phi_{e^-+e^+}^0(E_e).
\eqno(22)$$ $\eta=0.06$ is determined by the value of
$\Phi_{e^+}(E_e)$ at $E_e=20$ GeV, where there is no GC-effects. Our
predicted positron flux using Eqs. (16), (21) and (22) without other
extra parameters is shown in Fig. 15. The data are taken from
[53-56]. Now we can conclude that the increment of secondary
particles due to the GC-effects can explain the observed broken
power-law in electron and positron spectra, although the positron
background has uncertainty.

\newpage
\begin{center}
\section{Summary}
\end{center}

      A new discovery in QCD is that the gluons
in proton may converge to a state with a critical momentum at a high
energy range. This gluon condensation (GC) increases suddenly the
proton-proton or proton-nuclei cross sections. A natural suggestion
is that many observed, but uncomprehended excesses in cosmic ray
spectra origin from a common source-the GC. We find that the
GC-effects in $p-p(A)\rightarrow \pi^0\rightarrow \gamma$ sharply
break the gamma spectra. We present three possible broken power-laws
of eletron/postron spectra around $\sim 1$ TeV: a smooth break and
the sharp beaks with different sharpness using the GC-effects. The
relating excess of positron is also discussed. The GC is a new
phenomenon that is not yet known, it provides a new window to understand the high energy cosmic ray spectra.

\noindent {\bf Acknowledgments}: {\it Acknowledgments:} This work is
supported by the National Natural Science of China (No.11851303). We
thank X.R. Chen, L. Feng, D.B. Liu, P.M. Zhang and Y.P. Zhang for
useful discussions.

\end{document}